\documentclass[11pt]{article}
\usepackage{amsmath,amsthm,latexsym,amssymb,amsfonts}
\usepackage{graphicx, color, calc}
\usepackage[list=true, font=large, labelfont=bf,
labelformat=brace, position=top]{subcaption}
\usepackage{dsfont}

\makeatletter \@addtoreset{equation}{section} \makeatother

\title{On the volume simplicity constraint in the EPRL spin foam model}

\author{Benjamin Bahr$^1$, Vadim Belov${}^1$\\[5pt]
\small $^1$ II Institute for Theoretical Physics\\
\small University of Hamburg\\
\small  Luruper Chaussee 149\\
\small 22761 Hamburg, Germany
 }
\date{}

\begin{document}

\maketitle

\begin{abstract}
We propose a quantum version of the quadratic volume simplicity constraint for the EPRL spin foam model. It relies on a formula for the volume of 4-dimensional polyhedra, depending on its bivectors and the knotting class of its boundary graph. While this leads to no further condition for the 4-simplex, the constraint becomes non-trivial for more complicated boundary graphs. We show that, in the semi-classical limit of the hypercuboidal graph, the constraint turns into the geometricity condition observed recently by several authors. 
\end{abstract}
\section{Motivation}

The spin foam approach to quantum gravity rests on the formal equivalence of general relativity to a certain topological field theory, after the imposition of constraints. The topological theory in question is so-called BF theory, where the fields are a $Spin(4)$-connection $\omega$ on a $4d$ manifold $M$, and a bivector-valued 2-form, i.e.~a section $B$ in $\Omega^2(M)\otimes \bigwedge^2\mathbb{R}^4$, with the action
\begin{eqnarray}\label{Eq:BFTheory}
S_{BF}[\omega,B]\;=\;\int \text{tr}\left(B\wedge F[\omega],\right)
\end{eqnarray}%

\noindent where $F[\omega]$ is the curvature of $\omega$. Note we use $\bigwedge^2\mathbb{R}^4\simeq \mathfrak{so}(4)$. The constraints are imposed on the field $B$, which turns (\ref{Eq:BFTheory}) into the Holst action \cite{Holst:1995pc}, equivalent to GR on the classical level. The quantum theory is constructed by quantizing (\ref{Eq:BFTheory}) on a discretization of $M$, and imposing the simplicity constraints as (weak) operator equations on the boundary Hilbert space \cite{Barrett:1997gw, Engle:2007wy, Freidel:2007py, Baratin:2008du, Baratin:2011hp}.

They fall into three categories: diagonal-, cross- and volume-simplicity constraints. The original constructions of spin foam models within the context of loop quantum gravity, however, were dealing exclusively with discretizations of $M$ which were triangulations, i.e.~where the minimal building blocks of space-time are $4$-simplices. For this case, the volume-simplicity constraint follows from diagonal- and cross-simplicity, plus closure, which is one of the equations of motion. In essence, the volume simplicity constraint can be ignored in that case, and the resulting spin foam models that have been developed give a definition of the path integral of the theory. 

It was later that the constructed model was generalized, in a straightforward way, to discretizations which were not built on triangulations, where the minimal building block could, in principle, be any $4d$ polytope \cite{Kaminski:2009fm}. Although these models were well-defined, the fate of the volume-simplicity constraint remained obstructed in the more general case. 

In particular, in \cite{Bahr:2015gxa} it was observed that the generalized model contain non-geometric degrees of freedom in the case of the $4d$ hypercuboid. Later, these were also investigated in more general cases \cite{Bahr:2017eyi, Dona:2017dvf}. Their occurrence was interpreted as a lacking of implementation of the volume-simplicity constraint in this case, in \cite{Belov:2017who}.

The above findings suggest that additional care needs to be taken on the proper implementation of the missing set of constraints in the general case, thus bringing the volume part of simplicity under closer scrutiny. According to the modern perspective, diagonal- and cross-simplicity are treated linearly, using auxiliary 4d normals. Since the 4-volume constraint is decidedly quadratic in nature, this discrepancy may pose an additional question: which form of volume simplicity is an appropriate one to impose? One such alternative, formulated linearly, appeared in \cite{Gielen:2010cu}, and in \cite{Belov:2017who} we gave its dual reinterpretation, actually, as prescribing unambiguous 3-volume. 

In the following article, we investigate the quadratic volume-simplicity constraint on the classical and the quantum level.

\subsection*{Outline of the paper:} In section \ref{Sec:Setup} we will review the status of the discrete simplicity constraints in detail, on the classical level. In section \ref{Sec:BivectorGeometries} we take a look at so-called bivector geometries, and propose a conditions on the bivectors which serves as the discrete version of the volume simplicity constraint. In section \ref{Sec:Hypercuboid} we take a look at the case of the $4d$ hypercuboid, and discuss the new constraints in this case. From this discussion, we infer a proposal for an implementation of the discrete volume-simplicity constraint in section \ref{Sec:QuantumConstraints}. We close with a summary, and a discussion of open questions.

\section{Setup}\label{Sec:Setup}
In its Holst formulation, the Hilbert-Palatini action of GR can be written in terms of the spin connection $\omega$ and the vierbein $e$, with the Holst action
\begin{eqnarray}
S_{H}[\omega, e]\;=\;\int \left(*(e\wedge e)+\frac{1}{\gamma}e\wedge e\right)\wedge F[\omega],
\end{eqnarray}

\noindent where $*$ is the Hodge operator on the internal space, and $\gamma\neq 0,\pm1$ is the Barbero-Immirzi parameter, which does not influence the classical dynamics. 

The simplicity constraints can be reformulated in terms of $\Sigma$, which is defined by $B=\Sigma+\frac{1}{\gamma}*\Sigma$, or 
\begin{eqnarray}\label{Eq:Relation_B_Sigma}
\Sigma=\frac{\gamma^2}{\gamma^2-1}(B-\frac{1}{\gamma}*B),
\end{eqnarray}
\noindent such that $\Sigma\stackrel{!}{=}*(e\wedge e)$. They read
\begin{eqnarray}\label{Eq:OriginalConstraints}
\epsilon_{IJKL}\Sigma_{\mu\nu}^{IJ}\Sigma_{\sigma\rho}^{KL}\;=\;e\,\epsilon_{\mu\nu\sigma\rho}
\end{eqnarray}

\noindent in its quadratic form, where $e=\frac{1}{4!}\epsilon_{IJKL}\epsilon^{\mu\nu\sigma\rho}\Sigma_{\mu\nu}^{IJ}\Sigma_{\sigma\rho}^{KL}$. Classically these are imposed via Lagrange multipliers.%\footnote{ Using the quadratic form of the constraints (\ref{Eq:OriginalConstraints}) leads to more solutions than just $\Sigma=*(e\wedge e)$. This is remedied by going over to the linear version of the constraints, which we will do in the rest of the article. }

A main part of the spin foam quantization is to discretize space-time, and work with variables associated to elements on that discretization \cite{Engle:2007wy, Perez:2012wv}. One way to do this is to work on a polyhedral decomposition of $M$. Usually, the simplicity constraints are considered locally, i.e.~separately for each polytope $P$. The subpolyhedra of $P$ carry labels representing the fields. In the following, we characterize the polytope $P$ in terms of its boundary graph $\Gamma$. The vertices of $\Gamma$ correspond to $3$-dimensional polyhedra, and whenever two of these meet at a common face, $\Gamma$ has an edge $e$ between the two corresponding vertices. 

Since $\Sigma$ is a $2$-form it can be discretised by integrating it over the faces of $P$, corresponding to the edges $e$ in $\Gamma$.\footnote{There is inconsistent terminology among different authors, unfortunately. Usually, the terms \emph{edge} and \emph{vertex} are reserved for the $1$- and $0$-dimensional elements of the dual of the polyhedral decomposition, while \emph{link} and \emph{node} are reserved for the $1$- and $0$-dimensional elements of the boundary graph. However, since we will use the term \emph{Hopf link} later, and only consider a single polyhedron, we will use \emph{vertex} and \emph{edge} for the elements in boundary graphs, which is also in agreement with notions from graph theory.} So the $\Sigma$-field is realised as elements $\Sigma_e\in  \bigwedge^2\mathbb{R}^4$. For this integration, one needs to choose an orientation for the face, which is equivalent to choosing an orientation of the corresponding edge $e$ in $\Gamma$.

\begin{figure}
\begin{center}
\includegraphics[scale=1.0]{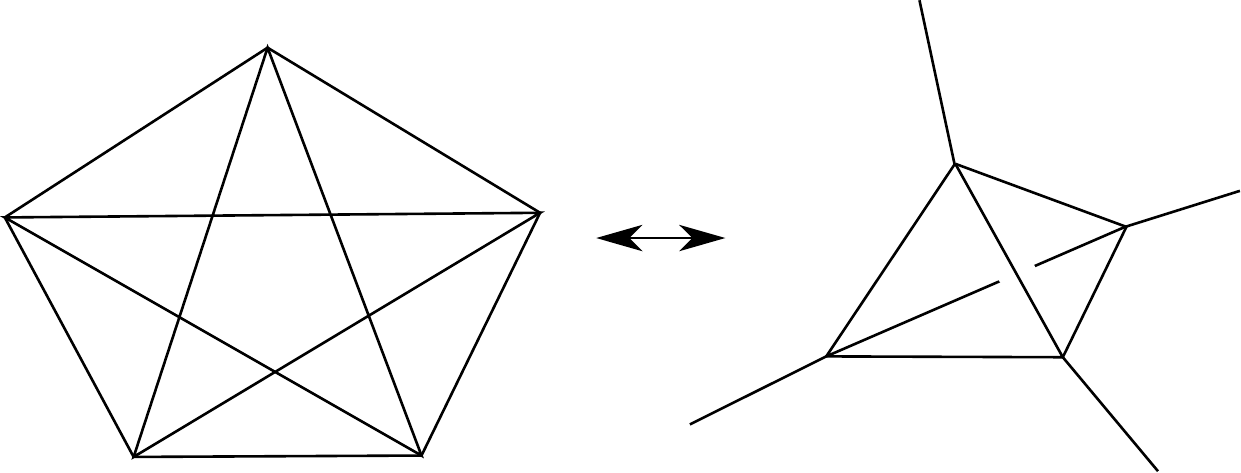}
\end{center}
\caption{A $4$-simplex, as polytope $P$. The boundary graph of $P$ is the complete graph in five vertices, and looks itself like a $4$-simplex. The graph can be rearranged to have only one crossing (additionally, we have moved one vertex to the infinitely far point, to make the image clearer).}\label{Fig:FourSimplex}
\end{figure}

In the early spin foam articles, the discretization was taken to be a triangulation, i.e.~where all polyhedra are $4$-simplices. For the case of a $4$-simplex, there are $10$ faces (i.e.~$10$ edges in the boundary graph, see figure \ref{Fig:FourSimplex}), and the simplicity constraints read

\begin{itemize}
\item For every edge $e$ one has that
\begin{eqnarray}\label{Eq:Simplicity_Diagonal}
\epsilon_{IJKL}\Sigma_e^{IJ}\Sigma_e^{KL}\;=\;0.
\end{eqnarray}
\item For every pair of edges $e$, $e'$ which are incident at the same vertex one has
\begin{eqnarray}\label{Eq:Simplicity_Cross}
\epsilon_{IJKL}\Sigma_{e}^{IJ}\Sigma_{e'}^{KL}\;=\;0. 
\end{eqnarray}  
\item For a pair of edges $e$, $e'$ which correspond to triangle which meet at exactly one point in the $4$-simplex, one has
\begin{eqnarray}\label{Eq:Simplicity_Volume}
\epsilon_{IJKL}\Sigma_{e}^{IJ}\Sigma_{e'}^{KL}\;=\;V.
\end{eqnarray}
\noindent  The latter constraint is to be read in a twofold way: both as the deﬁnition of the number $V_{ee'}$ (corresponding to a quantity on the l.h.s.~associated to a pair of edges), and as a requirement for it to be independent of a particular subset of edges chosen (since coming from the single expression (\ref{Eq:OriginalConstraints}) in the continuum). It is then interpreted as the 4-volume of a simplex.

This constraint is to be read as the definition of the number $V$, which is interpreted as the $4$-volume of the $4$-simplex. In particular, the number $V$ is required to be the same for every such pair of faces.
\end{itemize}

These constraints are called \emph{diagonal-simplicity}, \emph{cross-simplicity} and \emph{volume-simplicity}, respectively.

A major point in the definition of the EPRL- and FK-spin foam models \cite{Engle:2007wy, Freidel:2007py}, as well as the BO-model \cite{Baratin:2011hp}, is the replacement of the first two of these constraints, (\ref{Eq:Simplicity_Diagonal}) and (\ref{Eq:Simplicity_Cross}), with a linearized version. I.e.~for each vertex $v$ of the graph, it is required to exist a $4$-vector $N_v\in\mathbb{R}^4$ such that  
\begin{eqnarray}\label{Eq:Simplicity_Linear}
(N_v)^I(*\Sigma_e)_I{}^{J}\;=\;0\quad \text{for all }e\supset v.
\end{eqnarray}

\noindent It can be shown that (\ref{Eq:Simplicity_Linear}) implies (\ref{Eq:Simplicity_Diagonal}) and (\ref{Eq:Simplicity_Cross}). Furthermore, the volume simplicity constraint (\ref{Eq:Simplicity_Volume}) follows from (\ref{Eq:Simplicity_Linear}) plus the closure condition
\begin{eqnarray}\label{Eq:ClosureConstraint}
\sum_{e\supset v}[v,e]\,\Sigma_e\;=\;0
\end{eqnarray}
\noindent which is one of the equations of motion of $BF$ theory. Here $[v,e]=\pm 1$, depending on whether the edge $e$ is outgoing or incoming to $v$. In fact, (\ref{Eq:Simplicity_Linear}) is a slightly stronger condition than (\ref{Eq:Simplicity_Diagonal}), (\ref{Eq:Simplicity_Cross}), and restricts the possible solutions to $\Sigma_e=\pm*e_1\wedge e_2$. Hence, the linear simplicity constraint (\ref{Eq:Simplicity_Linear}) and closure condition (\ref{Eq:ClosureConstraint}) are used to define the models.

In \cite{Kaminski:2009fm}, the resulting EPRL model has been generalized to arbitrary $2$-complexes, in particular those dual to arbitrary polyhedral decompositions (but not limited to those). Since both linear simplicity and closure can be defined in this setting, the generalized model (sometimes dubbed EPRL-KKL) is a straightforward generalization. It should be noted however, that in the case of a polytope different from a $4$-simplex, the linear simplicity constraint and gauge invariance \emph{do not} imply the volume simplicity constraint (\ref{Eq:Simplicity_Volume})! 

One issue where this manifests itself is the large-$j$-asymptotics \cite{Barrett:2009gg}. For the $4$-simplex, it has been established that the asymptotic formula for  the vertex amplitude is exponentially suppressed in case the coherent boundary data does not describe the bivector geometry of a (possibly degenerate) $4$-simplex. This is a hint that (\ref{Eq:Simplicity_Linear}) and (\ref{Eq:ClosureConstraint}) are the correct way to constrain the d.o.f.~of BF theory to GR, in that precisely the geometric information remains. 

However, by now, there are several cases of other vertices known, in which this is not true any more \cite{Bahr:2015gxa,Dona:2017dvf}. In particular, the general case works on the level of $2$-complexes, which might not even be dual to a polyhedral decomposition at all. Therefore, it is a priori unclear how to define an analogue of (\ref{Eq:Simplicity_Volume}) at all, since the given formulation makes use of the $4$-simplex geometry explicitly, which might not even exist in the general case. 

In the following, we propose such a generalisation, for general graphs. 

\section{Bivector geometries and polyhedra}\label{Sec:BivectorGeometries}

While every polytope $P$ embedded in $\mathbb{R}^4$ uniquely determines the set of bivectors $\Sigma_e$ associated to its $2$-dimensional faces (corresponding to oriented edges $e$ in the boundary graph), it is an unsolved problem to reconstruct a polytope from its face bivectors. This construction lies at the heart of the simplicity constraints in the spin foam models for quantum gravity. The reason for this is that the Hilbert space vectors in the theory are the $SU(2)\times SU(2)$-spin network functions arising from the quantization of $BF$ theory, and the $\Sigma$-field of this theory is precisely what assigns bivectors $\Sigma_e\in\bigwedge^2\mathbb{R}^4$ to oriented edges $e$. The $\Sigma_e$ act naturally as operators on states $\Psi$, and thus the simplicity constraints in terms of the bivector operators are used to select a certain subset of $SU(2)\times SU(2)$ spin networks, corresponding to solutions to the simplicity constraints on the quantum level.

In the case of the $4$-simplex, the (classical, discrete) simplicity constraints (\ref{Eq:Simplicity_Diagonal}), (\ref{Eq:Simplicity_Cross}), and (\ref{Eq:ClosureConstraint}) are (apart from certain non-degeneracy conditions) precisely those constraints on a set of bivectors $\Sigma_e$ associated to faces of a $4$-simplex, which ensure that there exists a geometric $4$-simplex where the induced bivectors are precisely the given ones.

For a general state the situation is more complicated, since in general spin foam models there might not even be a polytope from which to take the bivectors. The only input we have is an oriented graph $\Gamma$, together with bivectors $\Sigma_e\in\bigwedge^2\mathbb{R}^4$ associated to the edges $e$.

We define a \emph{bivector geometry} to be a graph $\Gamma\subset S^3$ with oriented edges $e$, and an association of bivectors $\Sigma_e\in\bigwedge^2\mathbb{R}^4$ to the edges.\footnote{It is understood that any such geometry where the orientation of an edge $e$ is reversed and the corresponding $\Sigma_e$ is replaced by $-\Sigma_e$, defines that same bivector geometry.} We also demand the bivector geometry to satisfy the closure condition (\ref{Eq:ClosureConstraint}), as well as the linear simplicity constraint (\ref{Eq:Simplicity_Linear}). The latter implies (\ref{Eq:Simplicity_Diagonal}) and (\ref{Eq:Simplicity_Cross}).

\begin{figure}
\begin{center}
\includegraphics[scale=1.0]{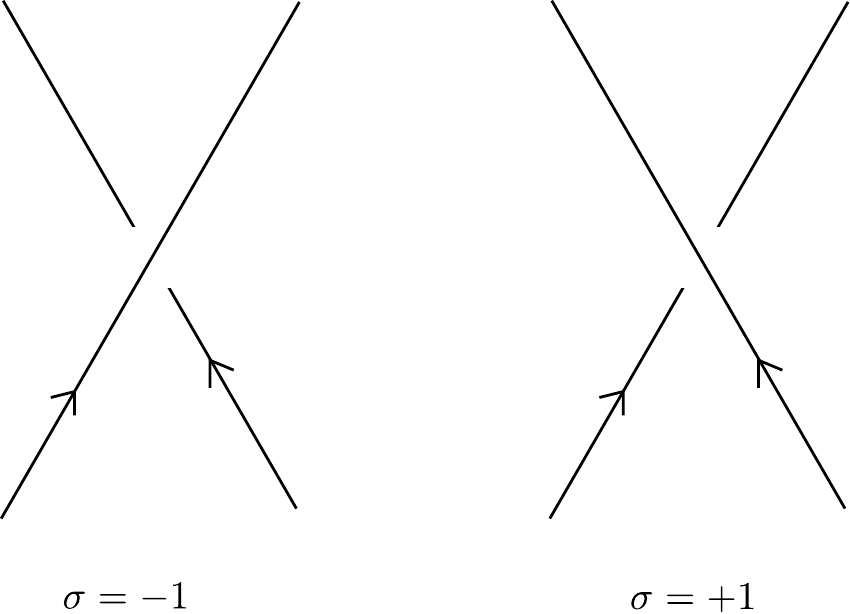}
\end{center}
\caption{With orientations, there are two types of crossings $C$ of edge in a graph $\Gamma$. They receive a crossing number $\sigma(C)=\pm 1$. }\label{Fig:TwoKnottingCases}
\end{figure}

\subsection{Hopf link volumes}

We define a set of numbers which rely on the embedding of $\Gamma\subset S^3$. In particular, we equip $S^3\subset\mathbb{R}^4$ with its standard orientation. By stereographic projection (with respect to a point which can be on $\Gamma$ itself), $\Gamma$ can be projected to $\mathbb{R}^3$, and from there to $\mathbb{R}^2$, with crossings.  Given the orientation of edges $e$ in $\Gamma$, there are two types of crossings, called \emph{positive} and \emph{negative}, depicted in figure \ref{Fig:TwoKnottingCases} . To these crossings $C$ we assign crossing numbers $\sigma(C)=\pm 1$, depending on their type.

For each crossing $C$, we define the corresponding crossing volume $V_C$ to be
\begin{eqnarray}\label{Eq:DefinitionCrossingVolume}
V_C\;:=\;\sigma(C)\;*\Big(\Sigma_1\wedge \Sigma_2\Big),
\end{eqnarray} 

\noindent where $\Sigma_{1,2}$ are the bivectors associated to the two edges within the crossing, and $*:\bigwedge^4\mathbb{R}^4\to
\mathbb{R}$ is the Hodge operator. We define the total $4$-volume of the bivector geometry to be
\begin{eqnarray}\label{Eq:DefinitionTotalVolume}
V\;:=\;\frac{1}{6}\sum_C V_C.
\end{eqnarray}

\begin{figure}
\begin{center}
\includegraphics[scale=0.5]{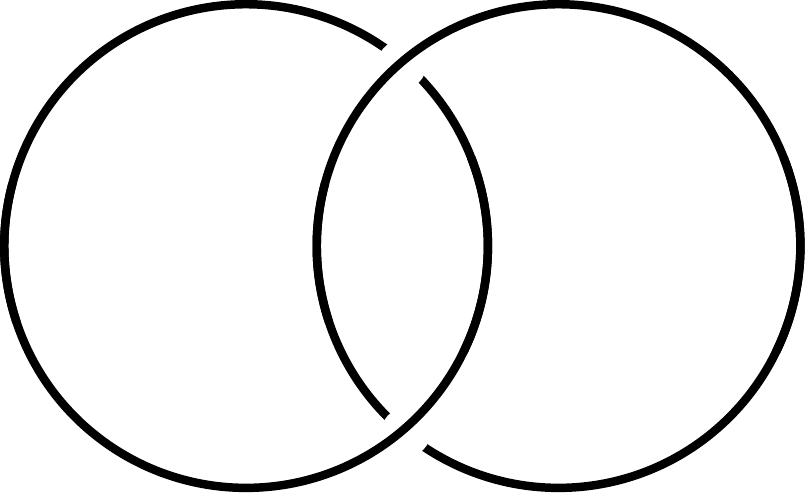}
\end{center}
\caption{A Hopf link, i.e.~two closed, linked curves, embedded in $S^3$.  }\label{Fig:HopfLink}
\end{figure}

\noindent Indeed, one can show that, in case the bivector geometry does come from a polytope $P\subset \mathbb{R}4$, $V$ is precisely the volume of $P$ \cite{Bahr:ToAppear01}. Furthermore, we define \emph{Hopf link volumes} the following way: A Hopf link $H$ in $\Gamma$ is defined to be a subset of edges in $\Gamma$ which form two linked (i.e.~non-intersecting) cycles  when embedded in $S^3$, see figure \ref{Fig:HopfLink}. Furthermore, they are not allowed to have any other crossing with any edge not in the Hopf link.

For a crossing $C$, we write $C\between H$, if it is between two edges of $H$. The Hopf-link volume $V_H$ associated to $H$ is then defined as
\begin{eqnarray}\label{Eq:HopfLinkVolume}
V_H\;:=\;\frac{1}{6}\sum_{C\between H}V_C.
\end{eqnarray}

\noindent It is not difficult so see that both $V$ and $V_H$ are independent of the embedding of $\Gamma$ to the plane with crossings. In other words, both are properties of the bivector geometry (and a choice of $H$) only. Also, both are clearly invariant under change of graph edge orientations. See \cite{Bahr:ToAppear01} for details. 

We say that the bivector geometry satisfies the \emph{Hopf link volume-simplicity constraint}, if $V_H$ is independent of the choice of Hopf link $H$ in $\Gamma$. In the following example, we will see that this is precisely the condition allowing for a reconstruction of the $4$-dimensional polytope from the bivector geometry.

\section{Case: the hypercuboid}\label{Sec:Hypercuboid}

In this section we consider the example of a bivector geometry which also occurs in the asymptotic analysis of the EPRL spin foam model \cite{Bahr:2015gxa}. It also plays a prominent role in renormalization computation of the model \cite{Bahr:2016hwc, Bahr:2017klw}. It is the prime example for a geometry in which diagonal- and cross-simplicity constraints are satisfied, but not necessarily volume simplicity. In particular, there does not, in general, exist a $4d$ polytope associated to it. However, the existence of such a $4d$ geometry can be formulated in terms of the Hopf volumes $V_H$, as we will show. 

\begin{figure}
\begin{center}
\includegraphics[scale=0.25]{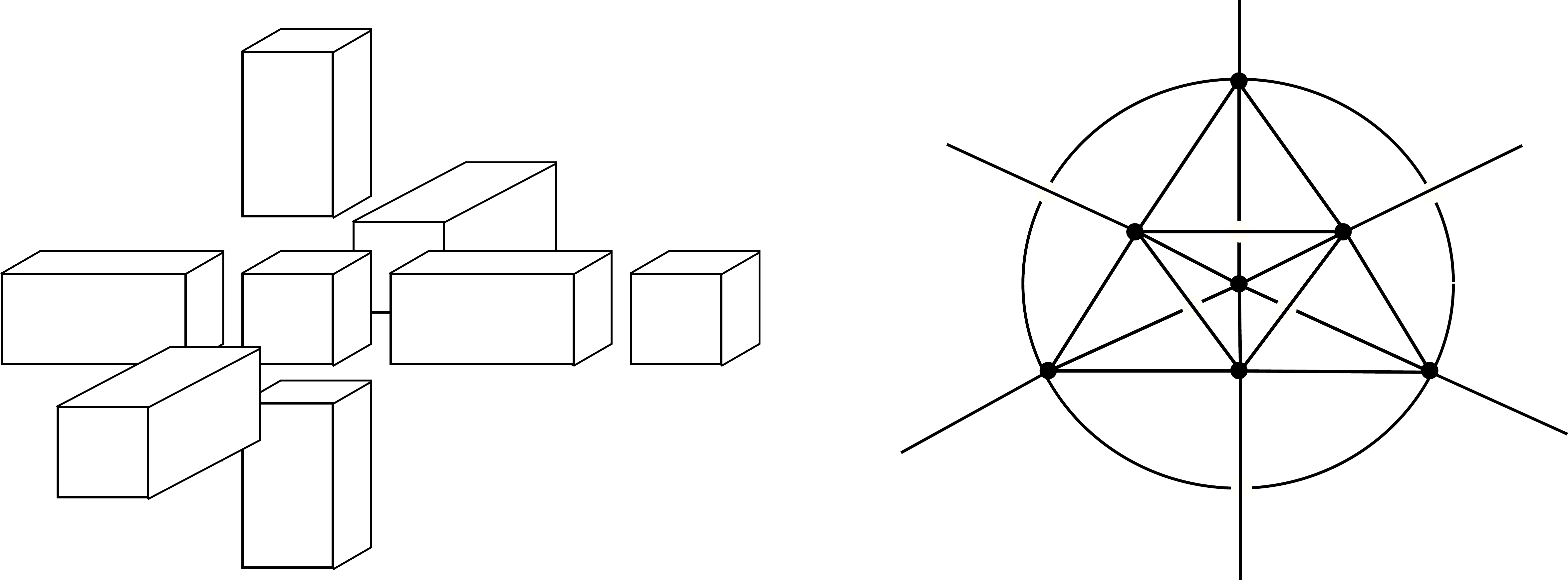}
\end{center}
\caption{The boundary of a hypercuboid, where every $3$-dimensional polyhedron is a cuboid. Its boundary graph, where each of the cuboids corresponds to one vertex. Vertex number 8 sits at the infinitely far point in $S^3$.}\label{Fig:HypercuboidGraph}
\end{figure}

\begin{figure}
\begin{center}
\includegraphics[scale=0.5]{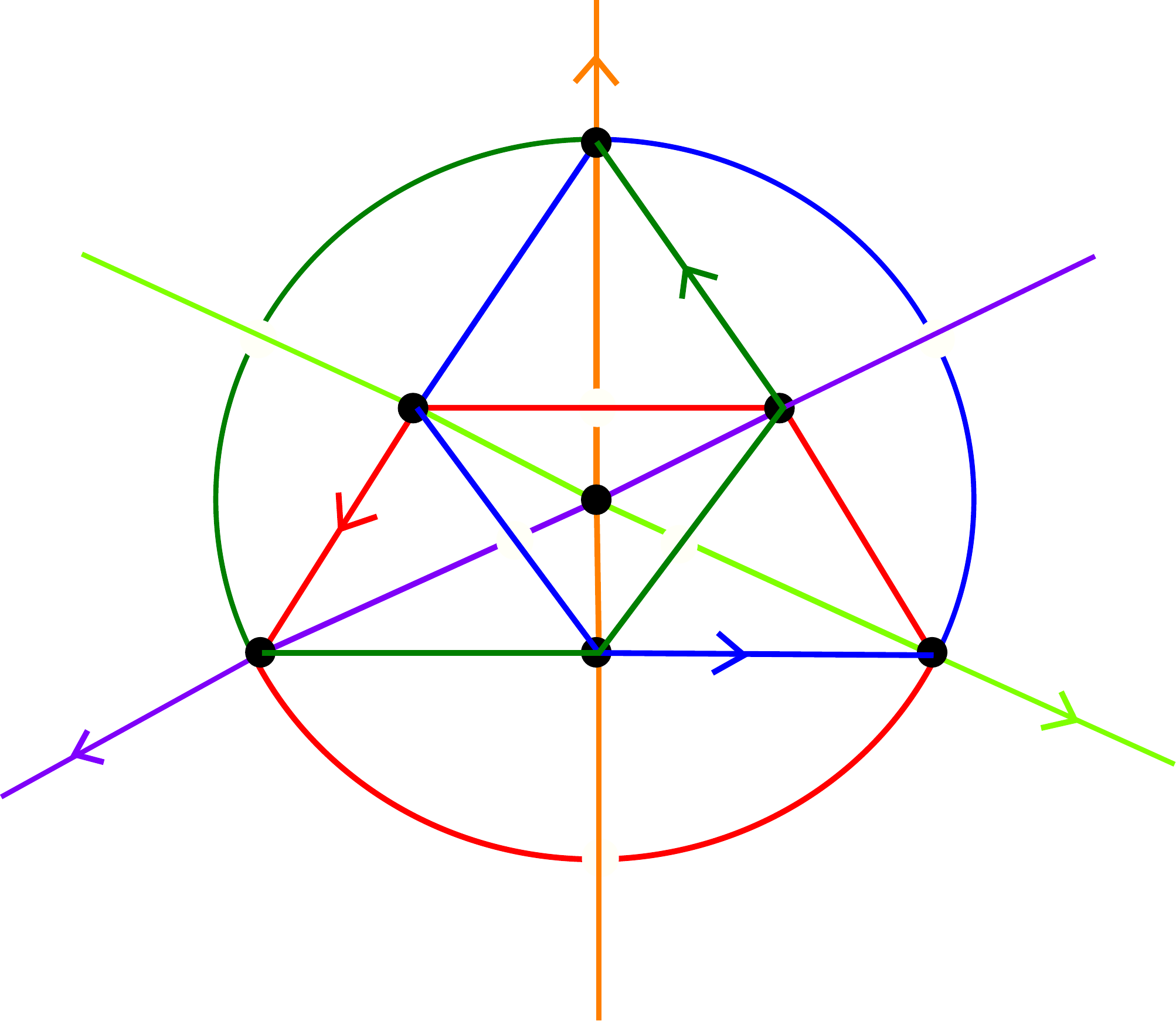}
\end{center}
\caption{The six independent loops in the hypercuboid graph $\Gamma$, colour-coded. Each such loop corresponds to one great circle on $S^3$. Two pairs form, respectively, one of the three independent Hopf links $H_1$, $H_2$, $H_3$ in $\Gamma$. The orientation of the edges in $\Gamma$ are such that each of the six crossings $C$ has $\sigma(C)=+1$.}\label{Fig:HopfLinkColour}
\end{figure}

The underlying graph $\Gamma$ is the dual boundary graph of a $4d$ hypercuboid (see figure \ref{Fig:HypercuboidGraph}). Note that, at this point, we only take the graph itself, but do not consider the hypercuboid as geometric, $4$-dimensional polytope. Rather, we consider a bivector geometry $\{\Sigma_e\}_e$ on $\Gamma$ such that, for each vertex of the graph, the incident bivectors form a $3d$ cuboid embedded in $\mathbb{R}^4$. Bivectors of such form are such that all $\Sigma_e$ lying on a great circle in $\Gamma$ coincide. Since there are six independent great circles in this graph (see figure \ref{Fig:HopfLinkColour}), this leaves us with the freedom of choosing six bivectors. They are as follows: 
\begin{eqnarray}\label{Eq:BivectorGeometryHypercuboid}
\Sigma_1\;&=&\;a_1\,*\big(e_y\wedge e_z\big),\quad \Sigma_2\;=\;a_2\,*\big(e_z\wedge e_x\big),\quad \Sigma_3\;=\;a_3\,*\big(e_x\wedge e_y\big),\\[5pt]\nonumber
\Sigma_4\;&=&\;a_4\,*\big(e_z\wedge e_t\big),\quad \Sigma_5\;=\;a_5*\,\big(e_t\wedge e_y\big),\quad \Sigma_6\;=\;a_6\,*\big(e_x\wedge e_t\big),
\end{eqnarray}

\noindent with areas $a_i>0$ and unit normal vectors $e_k\in\mathbb{R}^4$. It is straightforward to check that, for each vertex in the graph, the incident bivectors satisfy closure (\ref{Eq:ClosureConstraint}), as well as linear simplicity (\ref{Eq:Simplicity_Linear}). In particular, they also satisfy diagonal-, cross-simplicity, as well as the non-degeneracy conditions specified in \cite{Barrett:1997gw}.

However, the bivector geometry (\ref{Eq:BivectorGeometryHypercuboid}) does, in general, not follow from a $4d$ geometric polytope. While the restriction of (\ref{Eq:BivectorGeometryHypercuboid}) to each single vertex describes the  geometry of a $3d$ cuboid, these cuboids do not fit together in $\mathbb{R}^4$. While their touching faces are parallel and describe rectangles with coinciding areas, their shapes do not match (i.e.~they are rectangles with differing side lengths).

While the shape-matching problem is well-known in the $4$-simplex case, there the boundary states describing a non-shape-matching geometry are suppressed in the semiclassical asymptotics of the vertex amplitude. In the hypercuboid case, however, geometries as depicted above are not suppressed in that limit, as has been observed in \cite{Bahr:2015gxa, Dona:2017dvf}. This can be connected to the fact that, in case we do not deal with a $4$-simplex, linear simplicity and closure do \emph{not} imply volume simplicity. Hence, the set of allowed bivector geometries is not constrained enough, and also allows non-geometric configurations. 

However, the situation is different when we demand the bivector geometry to additionally satisfy the Hopf link volume-simplicity constraint, i.e.~if we demand that $V_H$ is independent of $H$.

In the case of the hypercuboid, there are essentially three different Hopf links (depicted in figure \ref{Fig:HopfLinkColour}). The respective Hopf-volumes are easily computed to be
\begin{eqnarray}
V_{H_1}\;=\;\frac{1}{3}a_1a_6,\qquad 
V_{H_2}\;=\;\frac{1}{3}a_2a_5,\qquad 
V_{H_3}\;=\;\frac{1}{3}a_3a_4.
\end{eqnarray}

\noindent Therefore, this version of the volume simplicity constraint reads 
\begin{eqnarray}\label{Eq:VolumeSimplicityHypercuboid}
a_1a_6=a_2a_5=a_3a_4.
\end{eqnarray}
\noindent It should be noted that these two conditions are precisely those that reduce the $6$-dimensional space of hypercuboidal bivector geometries (\ref{Eq:BivectorGeometryHypercuboid}) (given by the six $a_i$) to the $4$-dimensional subset of hypercuboidal geometries (given by four lengths of the edges of the hypercuboid): using (\ref{Eq:VolumeSimplicityHypercuboid}), one can easily show that the edge lengths of the $3d$ cuboids meeting at a common face derived from (\ref{Eq:BivectorGeometryHypercuboid}) agree, hence define four edge lengths, which in turn define a geometric hypercuboid in four dimensions.  

Thus, in the hypercuboid case, the Hopf link volume-simplicity constraint is precisely the right condition to allow for a reconstruction of the $4$-dimensional hypercuboid from the bivectors.

\section{Quantum constraints}\label{Sec:QuantumConstraints}

So far we have considered classical bivector geometries. In the following we will define a quantum version of the total volume and the Hopf link volumes, as well as discuss a version of the volume simplicity constraint derived from it.

The Hilbert space we use is the one from discrete BF theory, for gauge group $G=SU(2)\times SU(2)$. There is one Hilbert space $\mathcal{H}_\Gamma$ associated to the boundary of a $4d$ polytope with oriented boundary graph $\Gamma$. 
\begin{eqnarray}\label{Eq:BoundaryHilbertSpace}
\mathcal{H}_\Gamma\;=\;L^2\Big(G^E,d\mu_{\text{Haar}}\Big)^{\text{inv}}\;\simeq\;L^2\Big(G^E/G^V,d\mu\Big),
\end{eqnarray}

\noindent where $E$ is the number of edges in $\Gamma$, and $V$ the number of vertices. The invariance is with respect to the action of $k\in G^V$ on $g\in G^E$, i.e.~$g_e\,\to\,k_{s(e)}^{-1}g_e k_{t(e)}$, where $s(e)$ and $t(e)$ are source- and target vertex of the edge $e$. 

Due to the split of the gauge group $G=SU(2)\times SU(2)$ into right and left $SU(2)$, each state in (\ref{Eq:BoundaryHilbertSpace}) can be written as a linear combination of tensor products of two $SU(2)$ spin network functions $\Psi=\psi^+\otimes \psi^-$. The field $B_e=(\vec{b}_e^+,\vec{b}_e^-)$ acts as the respective left-invariant vector fields $(X^{+,I},X^{-,I})$, $I=1,2 ,3$, on $SU(2)$, i.e.
\begin{eqnarray}
\hat{b}_e^{+,\,I}\Psi\;=\;\Big(X_e^{+,I}\psi^+\Big)\otimes\psi^-,\qquad 
\hat{b}_e^{-,\,I}\Psi\;=\;\psi^+\otimes\Big(X_e^{-,I}\psi^-\Big).
\end{eqnarray}

\noindent Using the relation (\ref{Eq:Relation_B_Sigma}) between the $B$-field and $\Sigma$, we get
\begin{eqnarray}
\vec{\sigma}_e^\pm\;=\;\frac{\gamma}{1\pm\gamma}\vec{b}_e^\pm.
\end{eqnarray}

\noindent Since, for two bivectors $\Sigma_i\sim(\vec{\sigma}_i^+,\vec{\sigma}_i^-)$, with $i=1,2$, one has that
\begin{equation}
*(\Sigma_1\wedge \Sigma_2)\;=\;2\big(\vec{\sigma}_1^+\cdot\vec{\sigma}_2^+\,-\,\vec{\sigma}_1^-\cdot\vec{\sigma}_2^-\big),
\end{equation}

\noindent the crossing volume $V_C$ (\ref{Eq:DefinitionCrossingVolume}) of the two edges $e_{1,2}$, is quantized as
\begin{eqnarray}\label{Eq:CrossinVolumeOperator}
\hat{V}_C\;=\;2\sigma(C)\gamma^2\sum_{I=1}^3\left(\frac{1}{(1+\gamma)^2}X_{e_1}^{+,I}\otimes X_{e_2}^{+,I}\,-\,
\frac{1}{(1-\gamma)^2}X_{e_1}^{-,I}\otimes X_{e_2}^{-,I}
\right).
\end{eqnarray}

\noindent The expression (\ref{Eq:CrossinVolumeOperator}) has, up to factors, also appeared in \cite{Han:2011aa}, where it was introduced to construct an ad-hoc deformation of the EPRL spin foam model to include a cosmological constant. Using this, we similarly to (\ref{Eq:DefinitionTotalVolume}), (\ref{Eq:HopfLinkVolume}), define
\begin{eqnarray}
\hat{V}\;:=\;\frac{1}{6}\sum_C\hat{V}_C,\qquad \hat{V}_H\;:=\;\frac{1}{6}\sum_{C\between H}\hat{V}_C.
\end{eqnarray}

\noindent While it is easy to define operator analogues for total volume and Hopf link volume, the corresponding constraints on the quantum level are a bit tricky. There are two reasons for this. 

Firstly, it seems unlikely that the correct way of imposing the constraints is to impose it as strong or weak constraints on the state space. The reason  for that is that the volume constraint is  conceptually different from diagonal- and cross-simplicity (or linear simplicity, for that matter): The other constraints are decidedly \emph{kinematical}, since they involve conditions that can be formulated entirely in terms of single intertwiners of the $3d$ boundary data. Indeed, diagonal- and cross-simplicity (as well as closure) are used to guarantee that the intertwiners of the boundary state have an interpretation in terms of $3d$ polyhedra \cite{Livine:2007vk, Bianchi:2010gc}. The volume-simplicity constraint, however, relates bivectors on different, not necessarily neighbouring $3d$ polyhedra, which have a specific relation to each other depending on the $4d$ geometry. It is therefore arguably much more \emph{dynamical}. This is why the formulation of the constraints should not just be operator equations on the boundary Hilbert space, but should involve the dynamics of the model, i.e.~the amplitude.

Secondly, the so-called cosine problem makes the use of the direct amplitude difficult \cite{Barrett:2008wh}. Given of what we just said about the dynamical nature of the volume simplicity constraints, the most straightforward implementation of the Hopf-link volume-simplicity constraint would be to us the EPRL amplitude \cite{Engle:2007wy}
\begin{eqnarray}
\mathcal{A}(\Psi)\;:=\;\Psi(\mathds{1},\ldots,\mathds{1}),
\end{eqnarray}

\noindent and define
\begin{eqnarray}\label{Eq:HopfVolumesQuantum}
V_\Psi\;:=\;\mathcal{A}\big(\hat{V}\Psi\big),\qquad V_{H,\Psi}\;:=\;\mathcal{A}\big(\hat{V}_H\Psi\big),
\end{eqnarray}

\noindent demanding that 
\begin{eqnarray}\label{Eq:HopfLinkVolumeSimplicity}
V_H \text{ coincide for different Hopf links }H.
\end{eqnarray}

\noindent  However, the EPRL amplitude defines a dynamics in which both space-time orientations are being taken into account. This is easily seen in the semiclassical asymptotics \cite{Barrett:2009gg} of the amplitude $\mathcal{A}$, where not only $\exp(iS_{\text{EH}})$, but $\exp(iS_{\text{EH}})+\exp(-iS_{\text{EH}})=2\cos(S_{\text{EH}})$ appears. Since both of these contributions have $4$-dimensional volumes with opposite signs,  the semiclassical asymptotics of (\ref{Eq:HopfVolumesQuantum}) vanishes for the hypercuboid, as can be easily shown. This makes the comparison of (\ref{Eq:HopfVolumesQuantum}) for different $H$, as quantum version of the Hopf-link volume-simplicity constraints questionable.

There are several possible ways out of this: One would be to replace $\mathcal{A}$ with the so-called \emph{proper vertex} $\mathcal{A}^\text{pr}$ \cite{Engle:2011un}, which aims at defining the model to only include one of the two orientations. While its definition for the $4$-simplex is well-understood, the definition for arbitrary graphs is open, and relies on a definition of $4$-volume in that context. The definition of the constraint could become circular in that case. The semiclassical asymptotics of the proper vertex is, however, conjectured to be understood for all graphs, and in the hypercuboidal case leads to the right answer, as we will see below.

Alternatively, if one were to use the original EPRL amplitude, one could alter the definitions (\ref{Eq:HopfVolumesQuantum}), by replacing $\hat{V}_H$ by $\hat{V}_H^2$ or $|\hat{V}_H|$. Either of those should alleviate the cosine problem, since either expression is independent of the sign of space-time orientation.

In the semiclassical asymptotics, all three of these propositions give the same result. Let us consider the quantum cuboid states that have been introduced in \cite{Bahr:2015gxa}. These states have played a crucial role in the investigation of the critical behaviour of the EPRL model at small scales \cite{Bahr:2016hwc}. These states depend on six spins $j_i$, $i=1,\ldots, 6$ associated to the six great circles on the boundary, as depicted in figure \ref{Fig:HopfLinkColour}. The intertwiners are all given in terms of Livine-Speziale-intertwiners \cite{Livine:2007vk}
\begin{eqnarray}
\iota_{j_{a_1}j_{a_2}j_{a_3}}\;=\;\int_{SU(2)}dg\;g\triangleright\left[\bigotimes_{i=1}^3|j_{a_i},e_i\rangle\langle j_{a_i},e_i|\right],
\end{eqnarray}

\noindent which result in the boundary state 
\begin{eqnarray}
\Psi\;=\;\mathcal{P}\left[\bigotimes_{a=1}^8Y^\gamma\iota_a\right],
\end{eqnarray}

\noindent where the $\iota_a$ are the eight intertwiners corresponding to the eight cuboids in figure \ref{Fig:HypercuboidGraph}, $\mathcal{P}$ is the projector to the gauge-invariant subspace, and $Y^\gamma$ is the EPRL boosting map \cite{Engle:2007wy}. 

It has been shown that the large-$j$-asymptotic formula for the amplitude $\mathcal{A}(\Psi)$ in the case $\gamma\in(0,1)$ can be written as follows:
\begin{eqnarray}
\mathcal{A}(\Psi)\;\stackrel{j\to\lambda j,\lambda\to\infty}{\longrightarrow}\;\lambda^{-21}\left(\frac{1}{D}+\frac{2}{|D|}+\frac{1}{D^*}\right),
\end{eqnarray}

\noindent where $D$ is a complex polynomial of order $21$ in the $j_i$.

The proper vertex in that limit can be defined the following way: The asymptotics relies on the extended stationary phase approximation of the integral over $g^\pm_v\in (SU(2)\times SU(2))^{V}$ defining $\mathcal{A}$. For all known examples, there are at most two solutions $S_{1,2}$ for either $g^+_v$ and $g^-_v$, leading to four solutions in total. The proper vertex is defined by choosing $S_1$ for $g^+_v$ and $S_2$ for $g^-_v$ (see \cite{Barrett:2009gg, Engle:2011un, Engle:2015mra} for details). The proper vertex for the three different Hopf links $H_i$ in $\Gamma$  lead to
\begin{eqnarray}
\mathcal{A}^\text{pr}\big(\hat{V}_{H_1}\Psi\big)\;\longrightarrow\;\frac{\gamma^2}{6}j_1j_6\,\lambda^{-19}\left(\frac{1}{|D|}\right),
\end{eqnarray}

\noindent with $j_1j_6$ being replaced by $j_2j_5$ and $j_3j_4$ for $H_2$ and $H_3$, respectively. 

Using the square of the Hopf link volume operator and the EPRL amplitude results in 
\begin{eqnarray}
\mathcal{A}\big(\hat{V}_{H_1}^2\Psi\big)\;\longrightarrow\;\frac{\gamma^4}{36}\big(j_1j_6\big)^2\,\lambda^{-17}\left(\frac{1}{D}+\frac{1}{D^*}\right),
\end{eqnarray}

\noindent again with $j_1j_6$ being replaced by $j_2j_5$ and $j_3j_4$ for $H_2$ and $H_3$, respectively. Using $|\hat{V}_H|$ instead of $\hat{V}_H^2$ leads to a similar result. 

We see that in all of these cases, the Hopf-link volume-simplicity constraints can be formulated by demanding the corresponding quantum expressions to coincide for different Hopf links $H$ in the boundary graph. While they all agree in the asymptotic limit, in the regime of small spins they could, very well, differ. The question of which of the possibilities should be the right one, is still open. 

Nevertheless, we can see that in the semi-classical limit, the proper version of quantum condition turns into the classical condition (\ref{Eq:VolumeSimplicityHypercuboid}) on the boundary data $j_i,\iota_a$ of the coherent state $\Psi$, which in turn is precisely the geometricity condition discussed in \cite{Bahr:2015gxa, Dona:2017dvf, Belov:2017who}.

\subsection{Extension beyond the hypercuboidal case}

First we note that both operators (\ref{Eq:HopfVolumesQuantum}) are invariant under ambient isotopies, i.e.~they do not depend on which way the graph $\Gamma$ is projected to the plane \cite{Bahr:ToAppear01}. 

Let us comment about the $4$-simplex case. The boundary graph of the $4$-simplex is the complete graph in five vertices, which does not contain \emph{any} Hopf links. This is not hard to see, as any non-trivial cycle needs at least three vertices, so a Hopf link needs at least six in total, since the two circles in it are not allowed to intersect. So in this case the condition (\ref{Eq:HopfLinkVolumeSimplicity}) is empty, which is agreeing with the fact that already classically, the constraints (\ref{Eq:Simplicity_Diagonal}) and (\ref{Eq:Simplicity_Cross}) plus closure (\ref{Eq:ClosureConstraint}) imply the volume-simplicity constraint. 

On the other hand, the larger the graph $\Gamma$, the more independent conditions are given by (\ref{Eq:HopfLinkVolumeSimplicity}). This is in line with the fact that the classical condition (\ref{Eq:Simplicity_Volume}) gets more complicated for polyhedra with more faces. 

Still, it is an unsolved question whether, for arbitrary graphs $\Gamma$, conditions (\ref{Eq:HopfLinkVolumeSimplicity}) are sufficient to restrict the bivector geometry enough to capture the right discretised version of metric geometry. This is related to the question of whether closure, linear simplicity and Hopf-link-simplicity are enough to prove a reconstruction theorem analogous to the $4$-simplex case in \cite{Barrett:1997gw}. We feel that there is a very good chance for this to hold, since the Hopf-link constraint seems to capture the essential features of the $4$-volume in all cases we have looked at so far. Another hint might be the fact that the Hopf-link-volume $V_H$ could serve as a discretised version of the intersection form $Q$ of $4$-dimensional manifolds $M$. In its real formulation, it assigns a number $Q(\Sigma_1,\Sigma_2)$ to $\Sigma_i\in H^2(M)$, which, in the deRahm cohomology, are nothing but closed $2$-forms modulo translation symmetry. The expression in (\ref{Eq:DefinitionTotalVolume}) is then nothing but a discrete version of
\begin{eqnarray}
Q(\Sigma_1,\Sigma_2)=\int_M \Sigma_1\wedge \Sigma_2,
\end{eqnarray}

\noindent lifted to the context of $\Sigma_i\in H^2(M,\bigwedge^2\mathbb{R}^4)$ in the natural way. Due to the way in which the intersection form can also be evaluated by summing over the points of intersection of two  surfaces embedded in $P$, it might be possible to turn this into a sum over crossings of $1$-dimensional submanifolds in the boundary, i.e.~Hopf links in the boundary graph. It might therefore, indeed be enough to consider not the total volume, but only the Hopf link volume, to compute the volume of $P$. This is a point we will come back to in the future. 

Still, the question whether (\ref{Eq:HopfLinkVolumeSimplicity}) is sufficient for all graphs remains unanswered at this point. 

\section{Summary and conclusion}

In this article, we have proposed a conjectured quantum version of the quadratic volume simplicity constraint for the EPRL spin foam model. The goal is to provide an implementation of this constraint within the model, in order to restrict $BF$ theory to the Hilbert-Palatini action. In spin foam models the volume simplicity constraint is usually disregarded, since in the case of the $4$-simplex graph, it follows from the other simplicity constraints and the closure condition. This does no longer hold for polyhedra different from the $4$-simplex, which is why one could argue that the resulting models are not constrained enough. Indeed, it has been observed in several occasions that there are more than the usual geometric degrees of freedom present in the theory in the non-$4$-simplex case. The aim of introducing the quadratic volume constraint to the model is to select the right degrees of freedom also in the general case. 

We have constructed the constraint on the (discrete) classical level, as well as on the quantum level. Classically, the condition is translated to a discretized construction of the $4$-volume $V_H$ (\ref{Eq:HopfLinkVolume}) out of the bivectors. This construction relies on the choice of a Hopf link $H$ in the boundary graph. The constraint is then translated into the condition that the number is independent of the choice of $H$. Since the expression (\ref{Eq:HopfLinkVolume}) depends on the bivectors, which exist as operators on the boundary Hilbert space, there is a straightforward quantization. The quantum condition is then translated into the condition (\ref{Eq:HopfLinkVolumeSimplicity}) that the path-integral expectation value of the $\hat{V}_H$ are independent of $H$. The reason why this has to be formulated in this way, rather than as operator equation on the (kinematical) boundary Hilbert space, is that the volume-simplicity constraint is a statement about the $4$-dimensional geometry, and therefore inherently dynamical, in the sense of GR. 

In the case of the $4$-dimensional hypercuboid, we have shown that the imposition of (\ref{Eq:HopfLinkVolumeSimplicity}) is, in the large $j$ asymptotical limit, precisely satisfied by those states whose boundary data satisfies the geometricity constraints introduced in \cite{Bahr:2015gxa}. We conclude that the non-geometric degrees of freedom in the quantum cuboid case are a result of the insufficient imposition of the simplicity constraints in the EPRL model in this case, as had been conjectured in \cite{Belov:2017who}.

While the constraints we introduced can be defined straightforwardly for any kind of boundary graph, it is yet unknown whether (\ref{Eq:HopfLinkVolumeSimplicity}) is sufficient in those cases to properly impose the volume simplicity constraint. The question of general applicability rests on the following constraints (in increasing order of strength):
\begin{itemize}
\item Conjecture: For a general polytope $P$, the Hopf-link volumes $V_H$ are all independent of $H$.
\item Conjecture: For a general polytope $P$, the Hopf link volumes $V_H$ all satisfy $V_H=\frac{1}{3}V$, with $V$ being the total volume of $P$.
\item Conjecture: For any bivector geometry satisfying linear simplicity and closure, as well as all $V_H$ being equal, there is a (possibly degenerate) $4d$ polytope $P$ who induces this bivector geometry.
\end{itemize}

\noindent The latter conjecture would amount to a reconstruction theorem. To us, it seems not too improbable that these conjectures are true, since classically, there is a relation between the $4$-volume, and the intersection form $\int *(B\wedge B)$. The Hopf link volume can be regarded as a discretization of the intersection form, which makes the connection likely. Moreover, for a given convex polytope $P$ one can show that a sum over a set of Hopf links (which we called \emph{total volume}) is indeed a multiple of the volume of $P$. The volume simplicity constraint can therefore be regarded as going the other direction, i.e.~a condition on a bivector geometry to reconstruct $P$. Whether the Hopf link volume-simplicity constraint is sufficient for reconstruction (maybe with additional non-degeneracy conditions) of general convex polyhedra is, however, still open at this point. 

We should mention that, with the imposition of the constraints described in this article, the EPRL model becomes sensitive to the knotting of boundary graphs, which is not the case in its traditional  formulation, see e.g.~\cite{Bahr:2010my}.

Another possibility to be mentioned is to consider all constraints linearly, along the lines proposed in \cite{Baratin:2011hp, Belov:2017who}. This would probably necessitate first to enlarge the space of variables of the theory, and then quantize the respective topological model, which is elusive as of yet. (An adequate tool to capture the degrees of freedom of the extended configuration space into a single geometrical object seems to be the s.c.~Cartan connection, see e.g.~\cite{Wise:2009fu}.)

\section*{Acknowledgements}

This project was funded by the project BA 4966/1-1 of the German Research Foundation (DFG). The authors are indebted to Professor Nathan Bowler for discussions.

\bibliographystyle{jhep}
\bibliography{mybib}

\providecommand{\href}[2]{#2}\begingroup\raggedright\begin{thebibliography}{10}

\bibitem{Holst:1995pc}
S.~Holst, \emph{{Barbero's Hamiltonian derived from a generalized
  Hilbert-Palatini action}},
  \href{https://doi.org/10.1103/PhysRevD.53.5966}{\emph{Phys. Rev.} {\bfseries
  D53} (1996) 5966--5969},
  [\href{https://arxiv.org/abs/gr-qc/9511026}{{\ttfamily gr-qc/9511026}}].

\bibitem{Barrett:1997gw}
J.~W. Barrett and L.~Crane, \emph{{Relativistic spin networks and quantum
  gravity}}, \href{https://doi.org/10.1063/1.532254}{\emph{J. Math. Phys.}
  {\bfseries 39} (1998) 3296--3302},
  [\href{https://arxiv.org/abs/gr-qc/9709028}{{\ttfamily gr-qc/9709028}}].

\bibitem{Engle:2007wy}
J.~Engle, E.~Livine, R.~Pereira and C.~Rovelli, \emph{{LQG vertex with finite
  Immirzi parameter}},
  \href{https://doi.org/10.1016/j.nuclphysb.2008.02.018}{\emph{Nucl. Phys.}
  {\bfseries B799} (2008) 136--149},
  [\href{https://arxiv.org/abs/0711.0146}{{\ttfamily 0711.0146}}].

\bibitem{Freidel:2007py}
L.~Freidel and K.~Krasnov, \emph{{A New Spin Foam Model for 4d Gravity}},
  \href{https://doi.org/10.1088/0264-9381/25/12/125018}{\emph{Class. Quant.
  Grav.} {\bfseries 25} (2008) 125018},
  [\href{https://arxiv.org/abs/0708.1595}{{\ttfamily 0708.1595}}].

\bibitem{Baratin:2008du}
A.~Baratin, C.~Flori and T.~Thiemann, \emph{{The Holst Spin Foam Model via
  Cubulations}},
  \href{https://doi.org/10.1088/1367-2630/14/10/103054}{\emph{New J. Phys.}
  {\bfseries 14} (2012) 103054},
  [\href{https://arxiv.org/abs/0812.4055}{{\ttfamily 0812.4055}}].

\bibitem{Baratin:2011hp}
A.~Baratin and D.~Oriti, \emph{{Group field theory and simplicial gravity path
  integrals: A model for Holst-Plebanski gravity}},
  \href{https://doi.org/10.1103/PhysRevD.85.044003}{\emph{Phys. Rev.}
  {\bfseries D85} (2012) 044003},
  [\href{https://arxiv.org/abs/1111.5842}{{\ttfamily 1111.5842}}].

\bibitem{Kaminski:2009fm}
W.~Kaminski, M.~Kisielowski and J.~Lewandowski, \emph{{Spin-Foams for All Loop
  Quantum Gravity}}, \href{https://doi.org/10.1088/0264-9381/29/4/049502,
  10.1088/0264-9381/27/9/095006}{\emph{Class. Quant. Grav.} {\bfseries 27}
  (2010) 095006}, [\href{https://arxiv.org/abs/0909.0939}{{\ttfamily
  0909.0939}}].

\bibitem{Bahr:2015gxa}
B.~Bahr and S.~Steinhaus, \emph{{Investigation of the Spinfoam Path integral
  with Quantum Cuboid Intertwiners}},
  \href{https://doi.org/10.1103/PhysRevD.93.104029}{\emph{Phys. Rev.}
  {\bfseries D93} (2016) 104029},
  [\href{https://arxiv.org/abs/1508.07961}{{\ttfamily 1508.07961}}].

\bibitem{Bahr:2017eyi}
B.~Bahr, S.~Kloser and G.~Rabuffo, \emph{{Towards a Cosmological subsector of
  Spin Foam Quantum Gravity}},
  \href{https://doi.org/10.1103/PhysRevD.96.086009}{\emph{Phys. Rev.}
  {\bfseries D96} (2017) 086009},
  [\href{https://arxiv.org/abs/1704.03691}{{\ttfamily 1704.03691}}].

\bibitem{Dona:2017dvf}
P.~Dona, M.~Fanizza, G.~Sarno and S.~Speziale, \emph{{SU(2) graph invariants,
  Regge actions and polytopes}},
  \href{https://arxiv.org/abs/1708.01727}{{\ttfamily 1708.01727}}.

\bibitem{Belov:2017who}
V.~Belov, \emph{{Poincar\'e-Pleba\'nski formulation of GR and dual simplicity
  constraints}},  \href{https://arxiv.org/abs/1708.03182}{{\ttfamily
  1708.03182}}.

\bibitem{Gielen:2010cu}
S.~Gielen and D.~Oriti, \emph{{Classical general relativity as BF-Plebanski
  theory with linear constraints}},
  \href{https://doi.org/10.1088/0264-9381/27/18/185017}{\emph{Class. Quant.
  Grav.} {\bfseries 27} (2010) 185017},
  [\href{https://arxiv.org/abs/1004.5371}{{\ttfamily 1004.5371}}].

\bibitem{Perez:2012wv}
A.~Perez, \emph{{The Spin Foam Approach to Quantum Gravity}},
  \href{https://doi.org/10.12942/lrr-2013-3}{\emph{Living Rev. Rel.} {\bfseries
  16} (2013) 3}, [\href{https://arxiv.org/abs/1205.2019}{{\ttfamily
  1205.2019}}].

\bibitem{Barrett:2009gg}
J.~W. Barrett, R.~J. Dowdall, W.~J. Fairbairn, H.~Gomes and F.~Hellmann,
  \emph{{Asymptotic analysis of the EPRL four-simplex amplitude}},
  \href{https://doi.org/10.1063/1.3244218}{\emph{J. Math. Phys.} {\bfseries 50}
  (2009) 112504}, [\href{https://arxiv.org/abs/0902.1170}{{\ttfamily
  0902.1170}}].

\bibitem{Bahr:ToAppear01}
B.~Bahr, \emph{{Four-dimensional polyhedra and Hopf links}}, {\emph{to appear}
  }.

\bibitem{Bahr:2016hwc}
B.~Bahr and S.~Steinhaus, \emph{{Numerical evidence for a phase transition in
  4d spin foam quantum gravity}},
  \href{https://doi.org/10.1103/PhysRevLett.117.141302}{\emph{Phys. Rev. Lett.}
  {\bfseries 117} (2016) 141302},
  [\href{https://arxiv.org/abs/1605.07649}{{\ttfamily 1605.07649}}].

\bibitem{Bahr:2017klw}
B.~Bahr and S.~Steinhaus, \emph{{Hypercuboidal renormalization in spin foam
  quantum gravity}},
  \href{https://doi.org/10.1103/PhysRevD.95.126006}{\emph{Phys. Rev.}
  {\bfseries D95} (2017) 126006},
  [\href{https://arxiv.org/abs/1701.02311}{{\ttfamily 1701.02311}}].

\bibitem{Han:2011aa}
M.~Han, \emph{{Cosmological Constant in LQG Vertex Amplitude}},
  \href{https://doi.org/10.1103/PhysRevD.84.064010}{\emph{Phys. Rev.}
  {\bfseries D84} (2011) 064010},
  [\href{https://arxiv.org/abs/1105.2212}{{\ttfamily 1105.2212}}].

\bibitem{Livine:2007vk}
E.~R. Livine and S.~Speziale, \emph{{A New spinfoam vertex for quantum
  gravity}}, \href{https://doi.org/10.1103/PhysRevD.76.084028}{\emph{Phys.
  Rev.} {\bfseries D76} (2007) 084028},
  [\href{https://arxiv.org/abs/0705.0674}{{\ttfamily 0705.0674}}].

\bibitem{Bianchi:2010gc}
E.~Bianchi, P.~Dona and S.~Speziale, \emph{{Polyhedra in loop quantum
  gravity}}, \href{https://doi.org/10.1103/PhysRevD.83.044035}{\emph{Phys.
  Rev.} {\bfseries D83} (2011) 044035},
  [\href{https://arxiv.org/abs/1009.3402}{{\ttfamily 1009.3402}}].

\bibitem{Barrett:2008wh}
J.~W. Barrett and I.~Naish-Guzman, \emph{{The Ponzano-Regge model}},
  \href{https://doi.org/10.1088/0264-9381/26/15/155014}{\emph{Class. Quant.
  Grav.} {\bfseries 26} (2009) 155014},
  [\href{https://arxiv.org/abs/0803.3319}{{\ttfamily 0803.3319}}].

\bibitem{Engle:2011un}
J.~Engle, \emph{{Proposed proper Engle-Pereira-Rovelli-Livine vertex
  amplitude}}, \href{https://doi.org/10.1103/PhysRevD.87.084048}{\emph{Phys.
  Rev.} {\bfseries D87} (2013) 084048},
  [\href{https://arxiv.org/abs/1111.2865}{{\ttfamily 1111.2865}}].

\bibitem{Engle:2015mra}
J.~Engle and A.~Zipfel, \emph{{Lorentzian proper vertex amplitude: Classical
  analysis and quantum derivation}},
  \href{https://doi.org/10.1103/PhysRevD.94.064024}{\emph{Phys. Rev.}
  {\bfseries D94} (2016) 064024},
  [\href{https://arxiv.org/abs/1502.04640}{{\ttfamily 1502.04640}}].

\bibitem{Bahr:2010my}
B.~Bahr, \emph{{On knottings in the physical Hilbert space of LQG as given by
  the EPRL model}},
  \href{https://doi.org/10.1088/0264-9381/28/4/045002}{\emph{Class. Quant.
  Grav.} {\bfseries 28} (2011) 045002},
  [\href{https://arxiv.org/abs/1006.0700}{{\ttfamily 1006.0700}}].

\bibitem{Wise:2009fu}
D.~K. Wise, \emph{{Symmetric space Cartan connections and gravity in three and
  four dimensions}}, \href{https://doi.org/10.3842/SIGMA.2009.080}{\emph{SIGMA}
  {\bfseries 5} (2009) 080}, [\href{https://arxiv.org/abs/0904.1738}{{\ttfamily
  0904.1738}}].

\end{thebibliography}\endgroup

\end{document}